# Millimeter Wave Absorber for Secure Identification

Scott A. Skirlo, Jonathan T. Richard, Magued Nasr, Martin S. Heimbeck, John D. Joannopoulos,
Marin Soljacic, Henry O. Everitt, and Lawrence Domash

*Abstract*— **We demonstrate thin, flexible, metamaterial films with a strong, narrowband, polarization- and angle-insensitive absorption designed for wavelengths near one millimeter. These structures, fabricated by photolithography on a commercially available, copper-backed polyimide substrate, are nearly indistinguishable to the unaided human eye but can be easily observed by imaging at the resonance frequency of the film. We demonstrate that these patterns can be used to mark or barcode objects for secure identification with a terahertz imaging system.**

## I. Introduction

Millimeter wave (MMW) radiation in the 1 - 3 mm band penetrates dry dielectrics such as plastics, concrete, and fabric while being strongly absorbed by water and water vapor. This combination of characteristics may be exploited for numerous applications including short-range communications and radar, collision avoidance radar, non-destructive testing of materials and structures, security imaging, medical diagnosis, and spectroscopy [1-3]. Due to their limited range and high bandwidth, materials with interesting MMW properties also play a role in numerous security applications. One potential use could be to encode markings, signs, or barcodes on objects, such as railway cars or shipping containers that are invisible to the unaided human eye but can be read at high speed by narrowband scanning THz imaging systems operating at the correct frequency. Here, we demonstrate how such markings could be implemented with a wide angle, narrowband, absorptive resonance in a thin material and used to write a 2D spatial pattern readable only at a pre-specified MMW frequency by a terahertz imaging system.

Thin metamaterial 'perfect absorber' structures for the THz region have been explored in prior works [4-6]. One basic design family uses three layers; an opaque metal backplane layer, a dielectric spacer, and a plane of periodic metal islands whose lateral size is typically on the order of 1/4 the resonant wavelength. A "perfect absorber" design is reflective over a wide band of frequencies but strongly absorptive at one resonant frequency [7]. Such designs are intrinsically omni-directional, provided the islands are closely spaced. For a metamaterial resonant around 1 mm wavelength, the entire thickness of the three key layers can be less than 25 µm. Most previous demonstrations have been fabricated on semiconductor wafers, and some have been fabricated on flexible substrates [8,9]. Here we report a low cost approach where the substrate is a commercially available metal-coated polyimide. A single step of photolithography was sufficient to convert the substrate into a resonant metamaterial. To make larger areas, photolithography could be replaced by various electronic printing methods.

## II. Fabrication

Design and fabrication of metamaterial absorbers at microwave and terahertz frequencies using the three-layer backplane-dielectric spacer-metal array format have been extensively described [8,10-12]. In the layer containing an array of metal islands, a variety of patterns such as rings, split-rings, or cut wires can be used to create resonances. Tuning the geometrical parameters of a given design changes the effective permeability and permittivity of the metamaterial and the resonant frequency. The quality of the absorber is determined by how closely the impedance of the metamaterial matches free-space and by the magnitude of the imaginary parts of µ and  at resonance.

Our goal was to design and fabricate a flexible metamaterial film with strong absorption at approximately 290 GHz, building on a commercially available substrate. For ease of fabrication, we selected a pattern of square rings containing no extremely small features that might lead to manufacturing errors [8,13]. The copper-clad polyimide Pyralux LF7012R, made from a 12.7 µm film of Kapton with a 17.4 µm copper backing, was used to support these square rings. This composite has low loss, precise thickness, and has successfully been used by another group to fabricate metamaterial absorbers [9].

To design a panel with an omnidirectional, narrowband, polarization-independent response, the commercial electromagnetic solver Microwave CST was used, simulating the normal-incidence and large-angle absorption of the metamaterial for s and p polarizations. The simulation consisted of a single cell of the metamaterial with periodic boundary conditions for the in-plane directions. CST solved the transmission and reflection of the structure using a frequency domain form of the finite-integration technique,

Manuscript received April 19, 2016 *(Corresponding author: Henry Everitt)*

This work was supported the US Army Institute for Soldier Nanoscience at the Massachusetts Institute of Technology and Triton Systems Internal Research and Development Program 1500-197. The authors wish to thank John Blum for his contributions to alternative fabrication methodologies.

S. S., M. N. and J. R. were equal contributors.

S. S., J. J. and M. S. are with the Department of Physics, Massachusetts Institute of Technology, Cambridge, MA 02139.

J. R. is with IERUS Technologies, 2904 Westcorp Blvd Suite 210, Huntsville, AL 35805.

M. N. and L. D. are with Triton Systems Inc., 200 Turnpike Rd #2, Chelmsford, MA 01824.

M. H. and H. E. are with C.M. Bowden Lab, US Army Aviation and Missile RD&E Center, Redstone Arsenal, AL 35898 (henry.o.everitt.civ@mail.mil).

which is similar to the FDFD (finite-difference frequency-domain) method.

The resonance frequency of the absorber is sensitive to the lateral dimensions of the square rings, whereas the magnitude of the absorption depended mostly on the conductivity and thickness of the upper metal layer, since the spacer layer is fixed. This behavior is expected as the lateral dimensions of upper layer determine the effective inductance and capacitance of the resonator, while the conductivity and thickness affect resistance. The capacitive part of the resonator originates from the gaps between the neighboring metal rings, and the separation between the metal backplane and the rings [14]. We made the gaps between the rings large in the final design to minimize the error from this capacitance since we had the poorest tolerance for lateral dimensions. The inductive part of the resonance is controlled by the width and perimeter of the rings themselves.

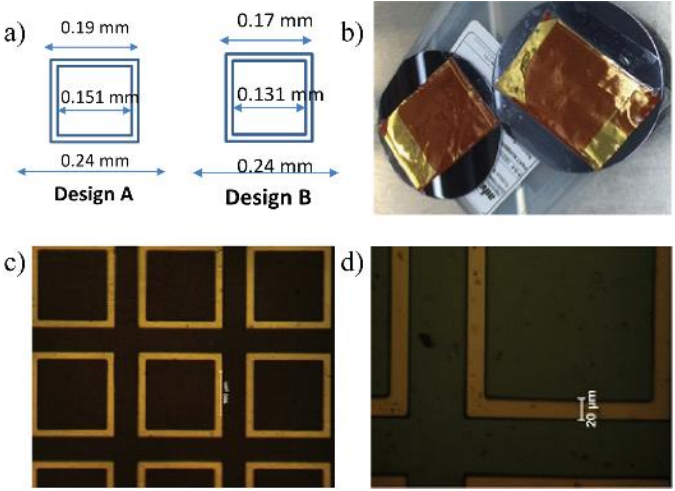

Fig. 1 (a) Two designs for different resonant frequencies. (b) Secure identification sample was made from strips of Design A metamaterial bonded onto unstructured Pyrolux. The pattern is nearly invisible to the naked eye. (c-d) Metamaterial film after photolithography, at two different magnifications.

Fig. 1a shows two of the unit cell layouts we investigated. Design A consisted of a hollow square about 190 µm in size, repeated on a 240 µm pitch, with a feature linewidth of about 20 µm, for a predicted center frequency of 285 GHz. In Design B we reduced the dimension for the hollow square to 170 µm, while keeping the pitch constant, to produce a higher resonant frequency about 320 GHz. We expect a higher resonance frequency for this design because the perimeter of the square ring is reduced, reducing the inductance of the ring and the capacitance with the backplane. From these arguments we roughly expect the new resonance frequency to be 285*0.19/0.17=318.5 GHz which matches the simulations fairly well. We simulated, fabricated and measured a number of such designs to investigate how varying these parameters affected the resonance frequency and absorption.

Fabrication of the hollow square pattern was by photolithography, sputtering 10 nm of Cr on the Kapton followed by 500 nm of Au. Figure 1c shows four images of sample films, at different magnifications. Calculated reflectivity spectra for Design A reveal that the resonance frequency is nearly independent of angle of incidence and polarization (Fig. 2).

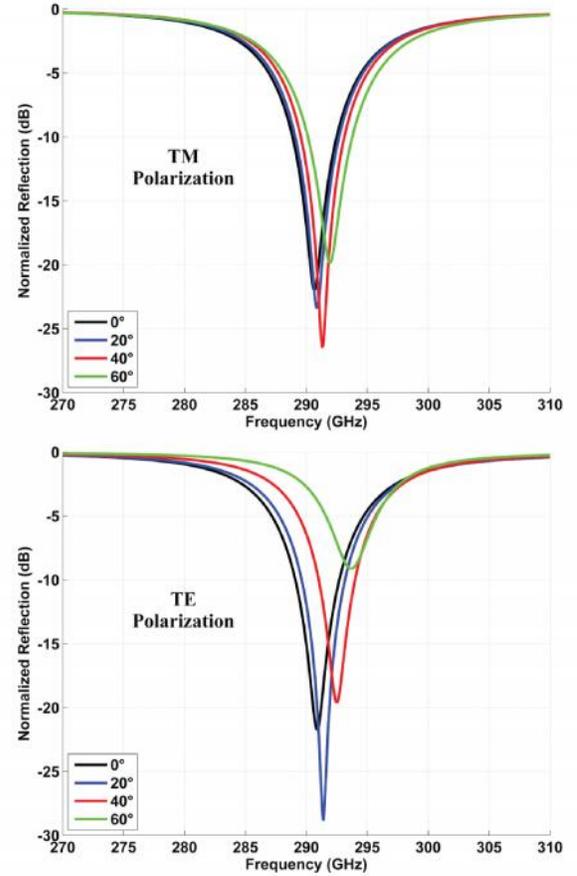

Fig. 2 Normalized reflection for Design A plotted as a function of incident angle and polarization, revealing insensitivity to these parameters.

### III. MEASUREMENT

The films were measured in a reflection geometry with an incident beam incident 40 degrees from normal using a Woollam terahertz spectroscopic ellipsometer [15]. The samples' spectral reflectivities were calculated by referencing the raw metamaterial signal to reflection measurements using a bulk aluminum mirror. The THz ellipsometer generated incident radiation using a frequency tunable backward wave oscillator and frequency multiplier, then recorded the reflected parallel and vertical polarized THz signal over a range of 220 – 330 GHz, as shown in Fig. 3. The measurements indicated an absorption peak at 292 GHz for Design A, slightly higher than the predicted frequency. Although no absorption peak was observed for Design B, an absorption tail was visible at the upper end of the frequency range, suggesting this peak was also slightly higher than the predicted frequency, just outside the spectral range of the measurement.

The small discrepancy between the calculated and measured resonant frequencies likely originates from the uncertainty associated with the dielectric constant of Kapton and manufacturing imperfections. The dielectric constant we used for Kapton in our original simulations was = 3.37+0.039i [16]. However, we found that the simulations showed the best agreement with experiment using a dielectric constant of 3.5+0.007i, a value consistent with other measurements performed at 60 GHz [17]. Small fabrication errors, such as the rounding of corners, are expected to broaden the absorption peak and shift its frequency.

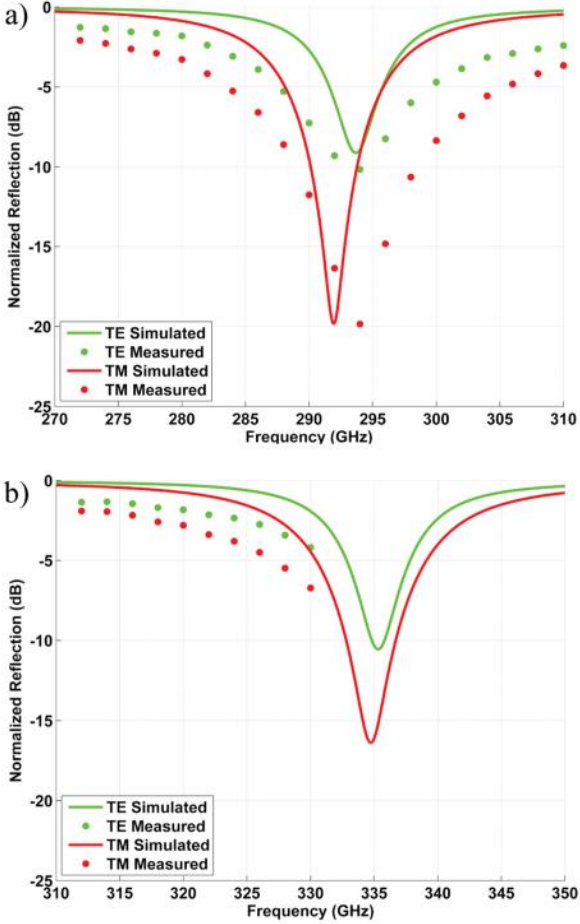

Fig. 3. The reflectance spectra of Designs A (a) and B (b), measured using the THz ellipsometer for 60° incidence and both linear polarizations, compared with the corresponding calculations.

Rozanov discovered a relationship between the absorption-bandwidth product and the absorber thickness, an observation that can be used to estimate how close our absorbers are to optimal [18,19]. Applying this to our normal incidence simulations for Design A, we find that the minimum thickness for any absorber design reaching the same absorbance performance is 7.6 µm. This is not much less than the 12.7 µm total thickness of our actual absorber, implying that the design is close to optimal.

### IV. SECURE IDENTIFICATION

To demonstrate the secure identification concept, 1 cm wide strips of the Design A material were bonded on a sheet of unstructured Pyrolux film to form a letter "A" about 15 cm high. The resulting sample was about 20 x 15 cm in area, and the metamaterial tape against the Pyrolux background was nearly invisible to the unaided eye (Fig. 1b). The strips were affixed with ordinary packaging tape, raising the refractive index of the medium surrounding the metamaterial and red shifting the resonance frequency from near 290 GHz to around 250 GHz. We can estimate the expected decrease assuming an index for tape of about 1.7. Since the increased index will increase the capacitance, we expect the new resonance frequency to be approximately $290/(1.7)^{1/4} = 254$ GHz which close to what we observe.

To observe this "invisible A", a terahertz confocal imager system was constructed from a Virginia Diodes G-band transceiver, which served as a frequency extender to an Agilent 5222A vector network analyzer (VNA). Specifically, the VNA provided the local oscillator (LO) and fundamental 10 – 20 GHz transmit signal (RF) for the Virginia Diodes 220 – 330 GHz transceiver. The LO and RF arrive at the transceiver input with a frequency offset of 279 MHz. The transceiver's internal hardware multiplies the LO and RF signals by a factor of 18 to produce an RF output of 220 – 330 GHz and an LO of 219.721 – 329.721 GHz. The LO is mixed with the reflected RF signal, producing a 270 MHz intermediate frequency (IF) that was processed by the VNA that recorded amplitude and phase information in I and Q format.

The transceiver used a collimating lens followed by a focal lens to interrogate a 2 mm diameter region of the sample, which is the smallest resolvable spot ("pixel"). At each pixel location, the frequency was swept from 220 to 330 GHz, and the receiver of the VNA referenced the measured reflected signal to the transmitted signal. Reflected signals were sampled at 1601 different frequencies, representing a frequency resolution of 68.7 MHz. These measurements were then combined into a single image using a raster scan of 100 vertical pixels by 75 horizontal pixels on a 2 mm pitch.

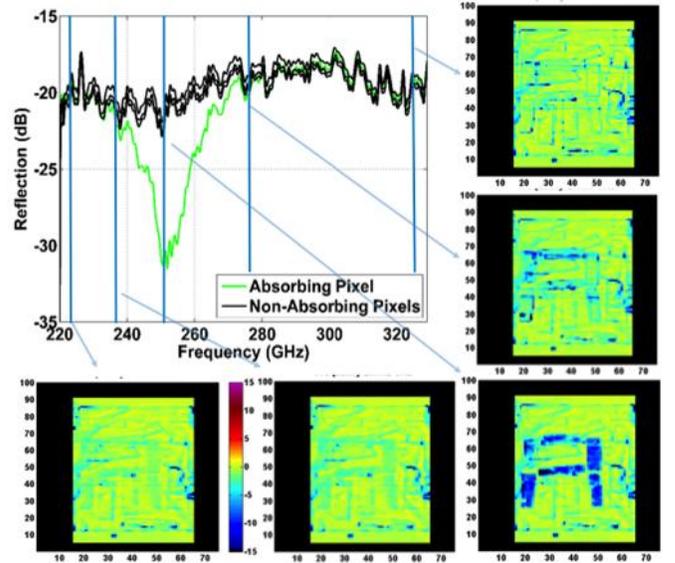

Fig. 4. Spectra of four pixels, one of which is located on the metamaterial region, and five normalized, range-gated 2D THz hyperplane images at (from left to right and bottom to top) 233, 237, 251, 278 and 326 GHz. The image of the letter A is seen most clearly near 251 GHz.

After the data was collected for each pixel, a wide range was applied through post-processing to isolate the object and plot the terahertz spectral hypercube. To accomplish this, an inverse fast Fourier transform (IFFT) was applied to the measured frequency-domain data of each pixel to generate a temporal signature from which the return range was obtained. The sample was range isolated by setting to -120 dB all range bins ±11 cm beyond the target (the noise floor itself spanned -70 to -90 dB, while the target reflection reached -30 dB). After applying a planar interpolation to compensate for non-normal placement of the sample, an FFT was applied to the remaining range domain data to produce a spectrum.

This spectrum represents a single pixel, a vertical column in the terahertz hypercube, four of which are plotted in Figure 4. Because of the contrast-enhancement and phase sensitivity of our imaging technique, the images show a pattern of changes in absorption that are visible over about a 50 GHz bandwidth, somewhat larger than the 3dB width of 30 GHz seen in the optical bench measurement of Figure 3. Horizontal planar slices through the THz hypercube produce spectrally resolved images, and five are also shown in Figure 4. A "movie" of these images, sweeping through the hyperplanes one frequency at a time, can be viewed in the supplement. The metamaterial tape is most clearly visible over a narrow (3.25 GHz FWHM) absorptive band near 250.2 GHz with >15 dB of attenuation. Averaging up to 15 hyperimages, thereby reducing spectral resolution to 1 GHz, did not significantly improve the image quality. This suggests excellent uniformity in size and shape of the constituent metamaterials throughout the panel.

## V. Discussion

The demonstration of secure identification imaging given here could be improved if, instead of a single frequency tape over an unstructured background, we were to use two or more different patterns with different metamaterial resonances, distributed over an area to form a spatial pattern representing characters or barcodes. These could be fabricated by single combined step of photolithography or printing.

Heterogeneous metamaterials with different resonant frequencies would then correspond to different spatial patterns at different spectral layers in the THz hypercube. By using two or more tapes with different resonant frequencies, the inscribed characters or message would change depending on the frequency of illumination. In this way, a movie of sorts could be encoded in the metamaterial structure that could be played simply by scanning the interrogating frequency. Within the moderate resolutions used here, larger areas of metamaterial film could easily be produced by screen printing with metallic inks, or other electronic printing methods.

It may also be possible to configure these materials as sensors by coupling the resonant frequencies of the metamaterials to local environmental conditions such as strain, moisture, or damage. Notice that the resonance frequency shifted approximately 40 GHz, more than a dozen linewidths, by the simple application of transparent packing tape over the metamaterial surface, indicating a strong sensitivity to the dielectric environment. Similarly, embedded metamaterial panels would represent a novel technique for structural monitoring of delamination, cracks in concrete, and incipient mechanical failure. Such an adaptation represents a step passive radio frequency identification (RFID) techniques by exploiting the superior resolution of MMW/THz probes to provide spatially-resolved status information from a target of interest. This opportunity is further enhanced by the comparatively strong attenuation provided by the atmosphere, limiting the problems posed by multi-site interference and constraining the propagation range of the interrogating MMW/THz beam. This technique will be best applied over short-ranges (< 1 km) and could provide significant when interrogating materials that may be sensitive to the much stronger electromagnetic signals typical of RFID interrogation scans.

## VI. Conclusion

We have demonstrated a simple method for making signage or barcodes which are nearly invisible to the naked eye or IR imagers but can be easily read by scanning or imaging at a specific frequency in the MMW range. Such materials appear manufacturable in large areas for application to security, the shipping industry, or structural monitoring.